# Nanoconfinement of Tetraphenylethylene in Zeolitic Metal-Organic Framework for Turn-on Mechanofluorochromic Stress Sensing

*Yang Zhang, Tao Xiong, Annika F. Möslein, Samraj Mollick, Vishal Kachwal, Arun Singh Babal, and Jin-Chong Tan\**

Multifunctional Materials & Composites (MMC) Laboratory, Department of Engineering Science, University of Oxford, Parks Road, Oxford OX1 3PJ, United Kingdom.

*Corresponding author
E-mail: jin-chong.tan@eng.ox.ac.uk

## Abstract

Mechanofluorochromic materials are of great significance for the fabrication of innovative sensors and optoelectronics. However, efficient mechanofluorochromic materials are rarely explored due to the deficiency of existing design strategies. Here, we demonstrate the incarceration of aggregation-induced emission (AIE) materials within metal-organic framework (MOF) single crystals to construct a composite system with turn-on mechanofluorochromism. A new type of AIE@MOF material was designed: integrating a zeolitic MOF (ZIF-71) and tetraphenylethylene (TPE, a topical AIE material) to generate a TPE@ZIF-71 system with exceptional turn-on type mechanofluorochromism. Using terahertz vibrational spectroscopy, we show the unique fluorochromism emanates from the enhanced nanoconfinement effect exerted by ZIF-71 host on TPE guest under pressure and its permanent fluorescence after stress release. Compared with pure TPE, we demonstrate the nanoconfinement in AIE@MOF not only changes the TPE's turn-off type sensing behavior to a turn-on type, but boosts the original sensitivity markedly by tenfold. Significantly, because ZIF-71 prevents the spontaneous recrystallization of TPE upon unloading, this allows TPE@ZIF-71 to record the stress history. This is the first demonstration of the Guest@MOF system combining the concepts of AIE and MOF; its promising properties and potential engineering applications will stimulate new directions pertaining to luminescent stress sensors and smart optics.

**Keywords:** Aggregation-induced emission (AIE); metal-organic framework (MOF); turn-on mechanofluorochromism; guest@MOF; photonic pressure sensor.

## 1. Introduction

Mechanofluorochromic material is a new type of smart material that changes its fluorescent color when subjected to an external mechanical force stimuli.[1] This kind of material has received extensive attention in the field of solid-state optics because of its potentially extensive applications in several advanced technologies,[2-4] such as fluorescence switches,[5] mechanosensors,[6-8] optoelectronics,[9] and data storage.[10]

Hitherto, there are only a few reported examples of materials that can exhibit efficient mechanofluorochromism by means of physical structural change.[1] The reason is that most of the existing fluorescent materials often have no emission in the solid-state due to the aggregation-caused quenching (ACQ) effect.[11] Even if some are able to show a weak mechanofluorochromic sensing behavior, they are essentially the "turn-off" types, which leads to difficulty in fabricating practical applications because turn-off type sensors are easily affected by external environmental factors, such as temperature and humidity, causing the intensity of the emission signal to decrease and interfere with the sensing accuracy.[12]

It has been proposed that one of the most effective ways to overcome the aforementioned obstacles is to prepare a "turn-on" type mechanofluorochromic material, which are not negatively affected by the ACQ effect.[13-16] Therefore, increasingly more researchers have turned their attention to study materials with aggregation-induced emission (AIE) property. The AIE materials, contrary to ACQ materials, have zero or only weak emission when in a diluted solution state but emit strong luminescence in the aggregated state,[17] and they may also possess mechanofluorochromic properties to some extent.[1]

In the field of AIE materials, one of the most well-known examples is tetraphenylethylene (TPE).[13, 18-22] Nonetheless, TPE only has a weak turn-off type sensing ability in the lower pressure range under 1.5 GPa.[23] Worse still, due to its spontaneous recrystallization, the stress-induced luminescence of TPE will immediately reverse to its (original) low-intensity state once the applied pressure is removed,[23, 24] which hinders measurement from being done *ex situ* (upon force removal). Many researchers have attempted to use organic chemistry methods to synthesize TPE derivatives for improving and optimizing its mechanofluorochromic properties,[1, 11, 25] but these methods have a limited success, and the complex synthesis implemented also limited its further development.

To address these challenges and promote the engineering application of TPE, herein we propose the use of metal-organic frameworks (MOFs) as a 'host' and TPE as a 'guest' to form a TPE@MOF system. Notably, MOFs have highly porous and extended crystalline structures, giving themselves the ability to capture, isolate and stabilize other emitter molecules, including bulky fluorophores.[26-32] We hypothesize that the nanoscale caging effect induced by MOF pores could modify or enhance the mechanofluorochromic performance of the pristine TPE molecules.

In this work, we report how TPE can be encapsulated into the zeolitic imidazolate framework-71 (ZIF-71) single crystal through an easy-to-implement one-pot synthetic method. Compared with the weak turn-off type sensing performance of pure TPE in the < 1.5 GPa pressure range, the introduction of ZIF-71 successfully allows the obtained TPE@ZIF-71 materials to exhibit turn-on sensing and enhances the sensitivity by 10 folds. Furthermore, the ZIF-71 solves the problem of spontaneous recrystallization in TPE, enabling the retainment of mechanofluorochromic behavior after the pressure is removed. Unlike the poor processability of other guest@MOF systems, we have successfully prepared turn-on type pressure sensing fibers and membranes with industrialization potential by leveraging TPE@ZIF-71/polymer composites. To the best of our knowledge, the TPE@ZIF-71 system is the first guest@MOF system to realize the turn-on type mechanofluorochromic sensing capability. This work has paved a new pathway for practical engineering applications of AIE-based mechanofluorochromic materials.

## 2. Synthesis and Structure of TPE@ZIF-71

The synthesis of TPE@ZIF-71, known as the one-pot synthesis method, is straightforward: by directly mixing solutions of zinc acetate, 4,5-dichloroimidazole (dcIm), and TPE at room temperature. The full details are given in the Experimental Section. Figure 1a and 1b show the morphology of the ZIF-71 and TPE@ZIF-71 crystals obtained, revealing the nominal size of the crystals to be ~200 nm. In addition, we performed powder X-ray diffraction (PXRD) characterization for the TPE, ZIF-71, and TPE@ZIF-71 samples. The PXRD spectra (Figure 1c) of TPE@ZIF-71 and ZIF-71 are indistinguishable from each other, and there is no Bragg peak belonging to TPE that also shows in the TPE@ZIF-71 spectrum. The patterns suggest that the crystal structure of ZIF-71 has formed normally, and the added TPE molecules do not affect the long-range periodicity of the ZIF-71 structure.

To ensure that the TPE is captured by the ZIF-71 pores instead of being attached to the surface, we conducted near-field infrared nanospectroscopy characterization. This technique combines scattering-type scanning near-field optical microscopy (s-SNOM) and nano-Fourier transform infrared (nanoFTIR) spectroscopy, which can perform detailed chemical and physical analysis on the surface of the nanocrystal to determine the presence/absence of guest material on the external surface of crystals.[33] The obtained s-SNOM optical phase images (Figure 1a and 1b) show that the surface composition of the ZIF-71 and TPE@ZIF-71 crystals are very uniform, and there are no clusters/regions of guest material on the surface. The nanoFTIR spectra of TPE, TPE@ZIF-71 and ZIF-71 (Figure 1d) reveal that the ZIF-71 and TPE@ZIF-71 are almost identical. Moreover, combined with the simulated infrared vibrations of TPE using density functional theory (DFT) (see Figure S1, Supporting Information), it can be found that the main peaks of TPE (1067.6 cm$^{-1}$, 1432 cm$^{-1}$, 1483 cm$^{-1}$, which correspond to the $C_{ph}$–H in-plane bending, and 1246 cm$^{-1}$, which corresponds to the C–$C_{ph}$ asymmetric stretching and the $C_{ph}$–H in-plane bending) are not detected in the IR spectrum of TPE@ZIF-71.

Similarly, using Fourier transform infrared spectroscopy with attenuated total reflection (ATR-FTIR) characterization (Figure S2) of the bulk polycrystalline sample, the traces of TPE molecules were not found in the TPE@ZIF-71 spectrum. The IR data indicate that the TPE molecules are not attached to the exterior of the ZIF-71 crystals. Subsequently, by employing thermogravimetric analysis (TGA, Figure S3), we found that the material contains only a low concentration (~0.24 wt.%) of TPE molecules. Additionally, according to the DFT simulation, the size of a TPE dimer is about 18.06 Å (including the *van der Waals* surface, see Figure S4); because the minimum distance inside the ZIF-71 pore is 16.58 Å, this means that most TPE guest can only exist as a monomer in the pores of the ZIF-71 host.

## 3. Fluorescent Properties of TPE@ZIF-71

To understand the photophysical properties of the resulting composite system, we performed fluorescence spectroscopy on the TPE (pristine guest), ZIF-71 (pristine host), and TPE@ZIF-71 samples (Figure 2). The excitation and emission spectra (Figure 2a and 2b) show that the luminescence of ZIF-71 and TPE@ZIF-71 are almost identical to each other. The excitation peak (300 – 350 nm) and emission peak (470 nm) pertaining to the pure TPE molecules do not appear in the TPE@ZIF-71 spectra. Therefore, we established that virtually all the fluorescence of TPE@ZIF-71 powder comes from the ZIF-71 itself. It is also found that the TPE is well dispersed in the ZIF-71, and the ZIF-71 pore does not effectively restrict the intramolecular vibrations of TPE in the absence of an external mechanical force stimulus (or pressure). Besides, ZIF-71 possesses an emission peak at 565 nm (Figure 2b), which is attributed to the ligand-metal charge transfer (LMCT) as identified by previous research.[34] Because the encapsulation of TPE materials does not lead to the disappearance of the LMCT peak, we may surmise that, there is initially no strong interaction between the TPE and ZIF-71.

Using the time-correlated single-photon-counting (TCSPC) technique to analyze the lifetime data of TPE@ZIF-71, ZIF-71, and TPE, we further verified the assumption made above. The lifetimes ($\tau$) of TPE@ZIF-71 and ZIF-71 shown in Figure 2c and Table S1 are the same, and the contributions ($c$) for each $\tau$ are also comparable, which proves that the fluorescence of TPE@ZIF-71 powder originates from the ZIF-71 alone, thereby confirming that: (i) TPE molecules are uniformly dispersed in ZIF-71, and (ii) there is no strong interaction between TPE and ZIF-71.

## 4. Mechanofluorochromism of TPE@ZIF-71

To study the mechanofluorochromism of the TPE@ZIF-71 system, we compressed it into pellets under different pressures using a hydraulic press. Figures 3a and 3b show the color of the pellets viewed under daylight and UV light (365 nm), respectively. Clearly, it can be seen with the naked eye that TPE@ZIF-71 exhibits a turn-on type mechanofluorochromic behavior even in the low-pressure range below 350 MPa. Conspicuously, the quantum yield (QY) of TPE@ZIF-71 pellets (Table S2) increases with increasing pelleting pressure. Moreover, the fluorescent property of the TPE@ZIF-71 pellets is maintained after pressure removal, which indicates that the nanoconfinement of ZIF-71 suppresses the spontaneous recrystallization of TPE molecules.

Then the excitation and emission spectra of these pellets were measured and analyzed (Figure 3). The excitation spectrum (Figure 3c) reveals that with the increase of pelleting pressure, the intensity of the peak belonging to TPE in the range of 300 – 350 nm gradually rises, and there is a small redshift, while the intensity of the peak belonging to ZIF-71 (400 – 450 nm) gradually decreases and a greater redshift appears. The emission diagram (Figure 3d) shows that the TPE@ZIF-71 pellets present a larger fluorescence intensity and peak wavelength under a higher pelleting pressure. The observed trend of increasing intensity and shifting wavelength also shows a certain degree of linear relationship with the applied nominal pressure (Figure 3e). From these phenomena, we inferred that as the pressure increases, the TPE molecules within the ZIF-71 pore are becoming more and more confined, resulting in a stronger caging effect and thus causing greater π−π interaction. This causes more restrictions to the intramolecular vibrations of TPE, giving a brighter red-shifted emission.[34, 35] In contrast, as a control experiment, a mechanically mixed powder of TPE and ZIF-71 was also prepared and subjected to the same pressure. It can be seen that, without the guest-host caging effect, this physically combined powder does not exhibit any mechanochromism under that same condition (Figure S5).

Detailed analysis of lifetime data further verifies our speculation. From Figure 3e and Table S3, it is clear that the lifetimes $\tau_1$, $\tau_2$, and $\tau_3$ of pellets rise with the incremental pressure, and the values become closer to the ones of pure TPE. In particular, when the pressure is above 259.95 MPa, the value of $\tau_3$ approximates to or even exceeds the $\tau_3$ of pure TPE. These findings prove that as the pressure increases, the emission of TPE starts to dominate, which results from the fact that the intramolecular vibrations of TPE are becoming more and more restricted. Moreover, Figure 3e and Table S3 also indicate that under a low pressure, the relationship between the contributions ($c$) of each lifetime is: $c_2 > c_1 > c_3$, and when the pressure increases, $c_3$ rises significantly, while $c_1$ decreases distinctly. The reason for this variation is that the fluorescence of TPE within the pores is turning increasingly dominant, because the contribution of pure ZIF-71 is $c_2 > c_1 > c_3$ and the contribution of pure TPE is $c_3 > c_2 > c_1$ (Figure 2c and Table S1). Hence, it follows that the data of lifetime further confirm our hypothesis on the relationship between the deformed structure and fluorescence property of TPE@ZIF-71.

Using synchrotron radiation Fourier-transform infrared spectroscopy (SR-FTIR) in Beamline B22 MIRIAM at the Diamond Light Source (Oxfordshire, UK), more evidence was obtained to substantiate the proposed mechanism. As shown in Figure 4 and Figure S6, we identified that: (i) the collective vibrational modes of TPE@ZIF-71 in the terahertz region (< 325 cm$^{-1}$, < 10 THz) broaden gradually with an increase of pelleting pressure; (ii) the vibrational peaks in the region of 490 – 590 cm$^{-1}$ (< 20 THz) decrease in their wavenumbers and progressively broaden as the pressure increases. From literature,[36] the former mode corresponds to the Zn–N bond stretching of ZIF-71, and the widening trend indicates that the framework is becoming structurally distorted or more amorphized under pressure; the latter spectral change is attributed to the dcIm ring mode vibrations, and the reduced wavenumbers represent the longer bond length, hence reflecting a weaker bond strength. The results support the notion that as the pressure increases, the TPE molecule begins to interact with the dcIm linker, and this intermolecular interaction becomes increasingly stronger with higher stress. The SR-FTIR spectra therefore confirm our proposed theory that the TPE experiences a more significant mechanically-induced caging effect as the pressure increases.

On the other hand, from the XRD spectra of pellets (Figure 3f), we could observe a progressive decline in intensity and sharpness of the XRD peaks, indicating that the crystalline structure of ZIF-71 is being amorphized with greater pressure. The evolving XRD patterns lead us to propose that, upon the collapse of the initially periodic framework, the TPE molecules in the adjacent pores could form new aggregates to help enhance the emission; such a mechanism explains the rise of intensity and redshift of the excitation and emission spectra. The rapture of the crystalline structure also results in the decrease and redshift of the excitation of ZIF-71. As the pressure progresses the framework is gradually amorphized, such a structural collapse could be triggered by a shear deformation,[37] thus reducing the distance between the linkers. Therefore, there will be more interaction between the discrete linkers or the remaining framework, causing the excitation peak to fall and redshift. The gradual disappearance of the 565 nm LMCT peak also proves our inference. The mechanism of structural change of TPE@ZIF-71 under pressure is illustrated in Scheme 1.

Through the comparison of other TPE@MOF composites, the proposed mechanofluorochromic mechanism is further validated. Meanwhile, the TPE@ZIF-71 system is also proven to have a better sensitivity than other TPE@MOF systems. Here we chose another two "stable" MOFs, UiO-67 (UiO: University of Oslo) and MIL-68(In) (MIL: Materials Institute Lavoisier), to encapsulate TPE. Figure S7 shows the mechanofluorochromic behavior of TPE@ZIF-71, TPE@UiO-67, and TPE@MIL-68(In) under the same pressure. It can be seen with the naked eye that TPE@ZIF-71 exhibits a more considerable peak intensity increase and wavelength shift as the pressure rises. The reason is closely related to the mechanical rigidity of the different frameworks. Due to the different organic linkers in the MOF structure, MIL-68(In) and UiO-67 are perceived to be more rigid than ZIF-71,[38-40] resulting in TPE experiencing a weaker caging effect inside MIL-68(In) and UiO-67, which explains why they both did not display as good of a sensing performance compared with TPE@ZIF-71. Concomitantly, this also proves the proposed turn-on sensing mechanism from another perspective.

Unlike other delicate guest@MOF systems,[35] TPE@ZIF-71 offers facile processability and unique turn-on luminescence under stress. This further motivates us to combine TPE@ZIF-71 with polyurethane (PU) and polyvinylidene difluoride (PVDF) matrices to prepare TPE@ZIF-71/PU fiber by electrospinning and TPE@ZIF-71/PVDF membrane by doctor-blade technique. Figure S8 shows that both the fibers and membranes maintain the excellent turn-on type mechanochromism and can realize fluorescence sensing under a compressive stress. To our best knowledge, this is also the first turn-on type mechanofluorochromic electrospun fiber and membrane to be demonstrated to date. The detailed performance of these proof-of-concept TPE@ZIF-71/polymer composites and their practical deployment warrant further investigations.

## 5. Conclusions

In summary, through a simple one-pot synthesis method, we successfully encapsulated the bulky TPE monomers as guest into the pores of the ZIF-71 host framework. The new TPE@ZIF-71 composite based on the Guest@MOF concept, for the first time, exhibits the turn-on mechanofluorochromic sensing behavior under stress. Furthermore, TPE@ZIF-71 can retain the results of the mechanofluorochromism after the pressure is removed. Utilizing far-infrared vibrational spectroscopy to probe the THz collective modes of TPE@ZIF-71, it was determined that this stress-induced response originates from the caging effect bestowed by MOF pores upon the TPE molecules, where (permanent) plastic deformation of the ZIF-71 framework constraints the intramolecular rotation and motion of TPE under stress/pressure. Compared with pure TPE and other TPE@MOF systems, the introduction of ZIF-71 not only reverses the turn-off type sensing of TPE (observed at pressure < 1.5 GPa)[23] to a turn-on type, but further boosts the sensitivity of TPE by 10 folds to detect a mechanical force stimulus (Table S4). The TPE@ZIF-71 system reported herein is the first study of its kind to combine AIE materials with stable MOFs, the high-sensitivity fluorescent properties and the potential engineering application of which have opened up a new way to design bespoke AIE materials by harnessing the nanoscale confinement effect.

**ACKNOWLEDGMENTS**

This work was supported by the EPSRC Impact Acceleration Account Award (EP/R511742/1) and the ERC Consolidator Grant (PROMOFS grant agreement 771575). We acknowledge the Diamond Light Source (Harwell, Oxford) for the award of beamtime SM25407; we thank Dr Mark Frogley for his assistance during the B22 MIRIAM beamline. We thank the Research Complex at Harwell (RCaH) for the provision of materials characterization facilities. We would like to acknowledge the use of the University of Oxford Advanced Research Computing (ARC) facility in carrying out this work (10.5281/zenodo.22558).

**AUTHOR CONTRIBUTIONS**

Conceptualization, Y.Z.; Methodology, Y.Z. and J.C.T.; Investigation, Y.Z.; Simulation, T.X.; s-SNOM and nanoFTIR, A.F.M. and Y.Z.; MIL-68(In) and UiO-67 Synthesis, S.M. and Y.Z.; Synchrotron Beamtime, Y.Z.,S.M. and A.F.M; Doctor Blade, S.M. and Y.Z.; Electrospinning,

V.K.; TGA, A.S.B and Y.Z.; Writing – Original Draft, Y.Z.; Writing – Review & Editing, Y.Z., T.X., A.F.M., S.M., V.K. and J.C.T.; Supervision, J.C.T.

**Experimental Section**

*Synthesis of TPE@ZIF-71*: 90 mL methanolic solution of 19.2 mmol 4,5-dichloroimidazole (dcIm), 10 mL tetrahydrofuran (THF) solution of 0.2 mmol tetraphenylethylene (TPE), and 90 mL methanolic solution of 4.8 mmol zinc acetate were quickly combined under stirring. After stirring for 24 hours, the sample was centrifuged at 8000 rpm for 10 minutes to remove excess reactants, and then washed 4 times (2 times with THF, 2 times with methanol). The crystals of TPE@ZIF-71 were separated from the suspension by centrifugation at 8000 rpm for 10 mins. The procedure for preparing ZIF-71 was the same, except no TPE was added.

*Synthesis of TPE@UiO-67*: 68 mL dimethylformamide (DMF) solution of 4.8 mmol biphenyl-4,4′- dicarboxylic acid (BPDC) plus triethylamine (NEt$_3$, 14.4 mmol) was combined with 10 mL THF solution of 0.2 mmol TPE. After the combination, 20 mL acetic acid solution of 4.8 mmol zirconium chloride was heated at 80 °C for one hour and then added into the mixture, producing a sol-like product. Subsequently, the product was heated at 120 °C for 24 hours and washed thoroughly 5 times (2 times with THF, 2 times with DMF, 1 time with methanol) to remove any excess guests adhered to the MOF surface. The nanocrystals were separated from the suspension by centrifugation at 8000 rpm for 10 mins.

*Synthesis of TPE@MIL-68(In)*: 9 mL DMF solution of 4.8 mmol benzenedicarboxylate (BDC) plus triethylamine (NEt$_3$, 9.6 mmol) was combined with 2 mL THF solution of 0.2 mmol TPE. After the combination, 6 mL DMF solution of 4.8 mmol indium nitrate was immediately added into the mixture. Then the product was washed thoroughly 5 times (2 times with THF, 2 times with DMF, 1 time with methanol) to remove the guests adhered to the MOF surface. The nanocrystals of TPE@MIL-68(In) were separated from the suspension by centrifugation at 8000 rpm for 10 mins.

*Sample Preparation for Fluorescence Characterization*: TPE suspension was prepared by diluting 0.1 M TPE THF solution in water (the volume ratio of water to THF was 99:1). The pellets for mechanofluorochromism study were made by using a manual hydraulic press (Specac Atlas) with a 1.2 cm diameter die under compressive forces of 1, 2, 3, and 4 tones.

**Electrospinning:** The solution for electrospinning was prepared by dissolving 13.7 g of polyurethane (PU) in 86.3 g of *N*, *N*-dimethylformamide (DMF), yeilding 13.7 wt.% of PU in DMF, and afterward, 100 mg of TPE@ZIF-71 powder was added into a 10 g of as-prepared solution. The mixed solution was loaded into a 5 mL glass syringe with a G-19 needle emitter (nozzle) connected and dispensed utilizing an automatic syringe pump. The infusion rate was set to 0.25 mL/h. A high voltage of 11.5 kV was applied between the nozzle and the base collector set 18 cm vertically apart.

**Doctor Blade:** The polymer solution for doctor-blade membrane fabrication was prepared by dissolving 13.7 g HSV900 polyvinylidene fluoride (PVDF) powder in 86.3 g dimethylacetamide (DMA) to yield 13.7 wt.% PVDF in DMA. Then, 100 mg of TPE@ZIF-71 powder was added into the 10 g of 13.7 wt.% PVDF solution. The membranes were deposited using an automatic doctor-blade apparatus (MTI Corporation, MSK-AFA-II). The thickness was set to 0.95 mm, and the coating speed was set to 10 mm/s.

*Materials Characterization*: The morphologies, optical phase images, and nanoFTIR results were examined using the neaSNOM instrument (neaspec GmbH) based on a tapping-mode AFM where a platinum-coated tip (cantilever resonance frequency 250 kHz, nominal tip radius ~20 nm) was illuminated by a broadband infrared laser. To suppress background contributions, the signal was modulated at the third harmonic of the tip frequency for optical phase images, and at the second harmonic for nanoFTIR absorption spectra. Each spectrum was obtained from averaging over 14 individual measurements with an integration time of 14 seconds, and subsequently normalized to the spectrum of the silicon substrate. The PXRD pattern was recorded using a Rigaku MiniFlex with a Cu Kα source (1.541 Å). Steady-state fluorescent spectra, lifetime, and QY were recorded employing the FS-5 spectrofluorometer (Edinburgh Instruments). For lifetime TCSPC measurement, a 362.5 nm laser was used and the stop condition was set to be at 10,000 counts. FTIR results were recorded by using a Nicolet iS10 FTIR spectrometer. TGA was performed using a TGA-Q50 machine (TA Instruments) equipped with a platinum sample holder under an $N_2$ inert atmosphere at a heating rate of 10 °C/min from 30 to 800 °C.

*Synchrotron Radiation Infrared Spectroscopy*: High-resolution infrared (IR) vibrational spectra of all compounds were recorded at the Multimode InfraRed Imaging and Microspectroscopy (MIRIAM) Beamline B22 at the Diamond Light Source synchrotron. IR spectroscopy was performed in vacuum *via* a Bruker Vertex 80 V Fourier Transform IR (FTIR) with an Attenuated Total Reflection (ATR) accessory (Bruker Optics, Germany). The mid-IR spectra were collected using a standard DLaDTGS detector. For the far-infrared spectral range below 700 cm$^{-1}$, a bolometer cooled by liquid helium was used for the detection of terahertz signals. All spectra were acquired with a resolution of 4 cm$^{-1}$ and a scanner velocity of 20 kHz.

***Simulation of TPE molecules***: Geometry optimization of the electronic ground state of TPE was performed using density functional theory (DFT) at the B3LYP/6-311G* level of theory,[41-45] followed by normal mode analysis to confirm that the stationary structure is a local minimum, using the quantum chemical package Gaussian 16.[46] The vibrational frequencies obtained were scaled by an empirical factor of 0.97.[47] Also, efforts to obtain the simplest form of aggregates, i.e. dimers, were made to evaluate the possibility of encapsulation of aggregates by the MOF. For the simulations for dimers, the DFT-D3 version of Grimme et al.[48] was used to account for the dispersion correction.

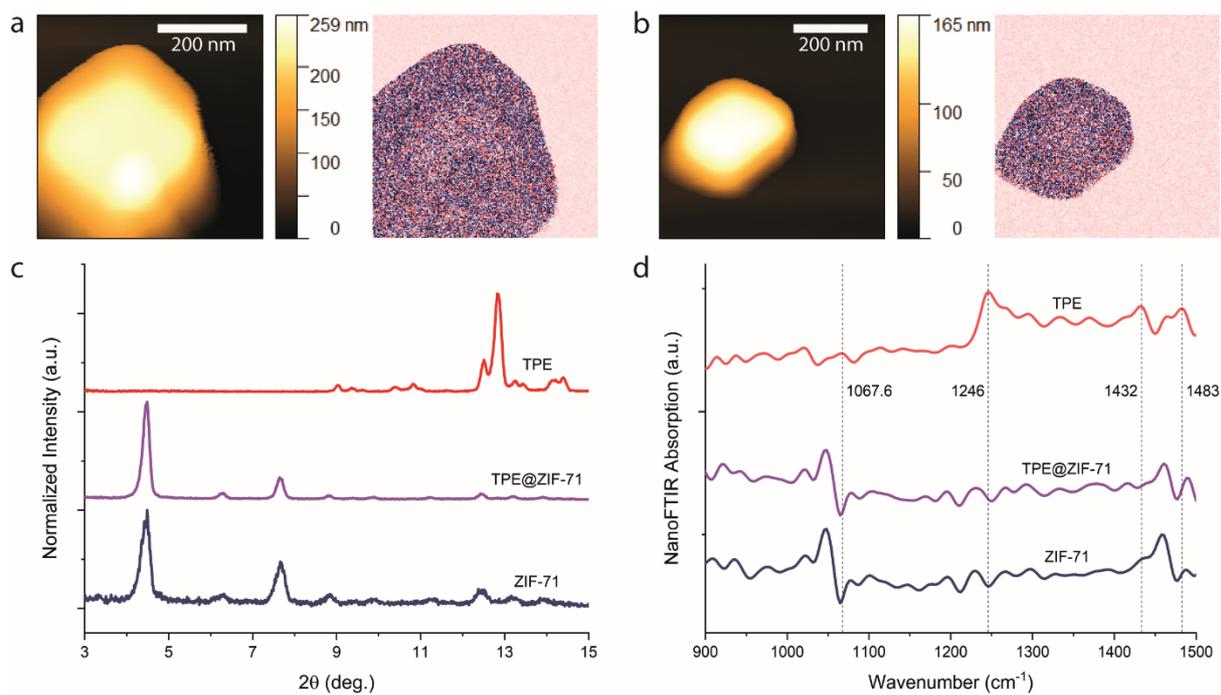

**Figure 1.** a) and b) Atomic force microscopy (AFM) images (left) and optical phase images (right) of ZIF-71 and TPE@ZIF-71, respectively. c) PXRD patterns of the TPE, TPE@ZIF-71, and ZIF-71 powder. d) Near-field IR absorption spectra of TPE, TPE@ZIF-71, and ZIF-71 crystals.

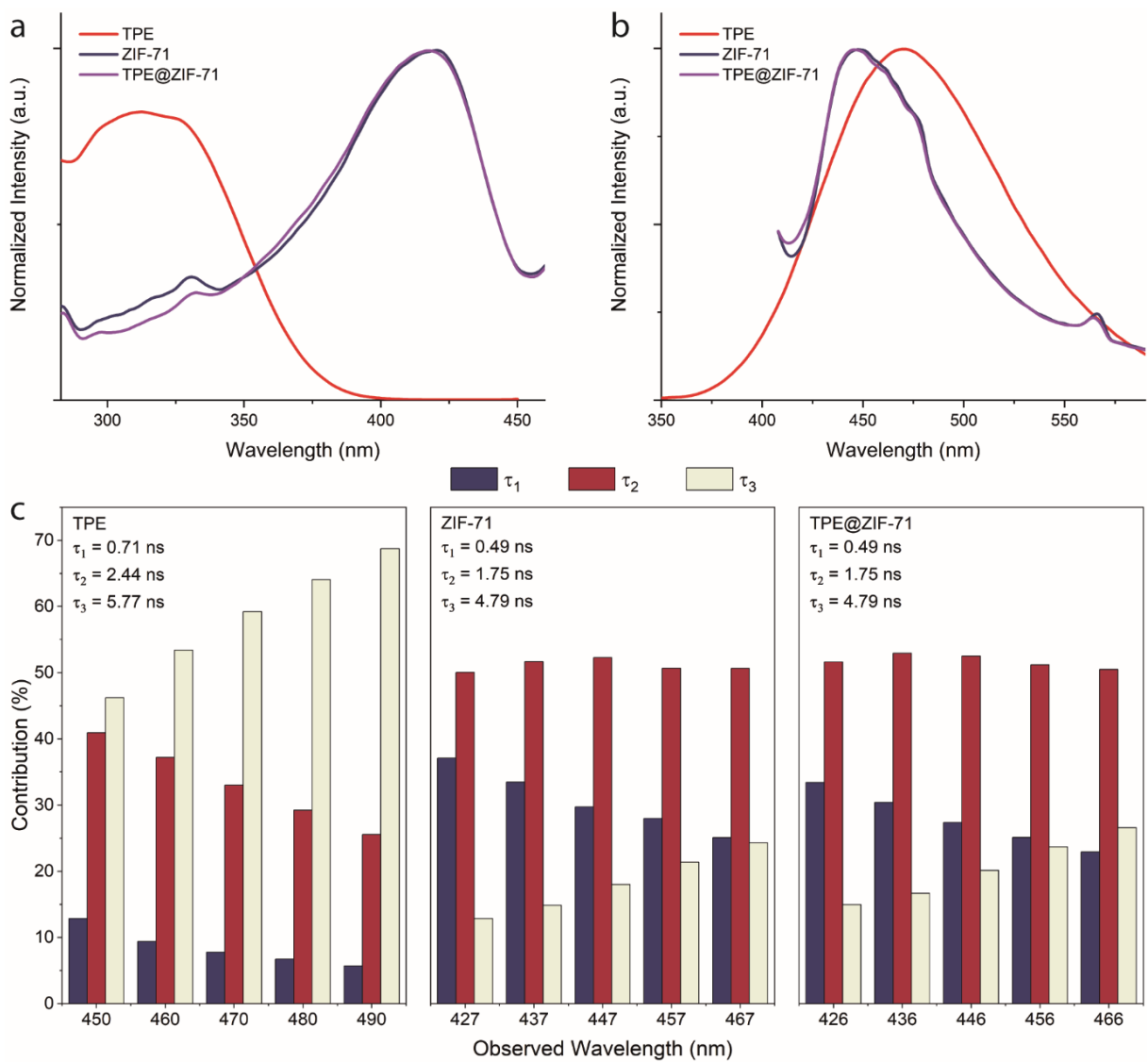

**Figure 2.** a) Normalized excitation spectra, b) emission spectra and c) lifetime data of TPE suspension (THF-water mixture containing 99% volume fraction of water), ZIF-71, and TPE@ZIF-71 powder. Time constants ($\tau_i$) and fractional contributions from fluorescent lifetime measurements.

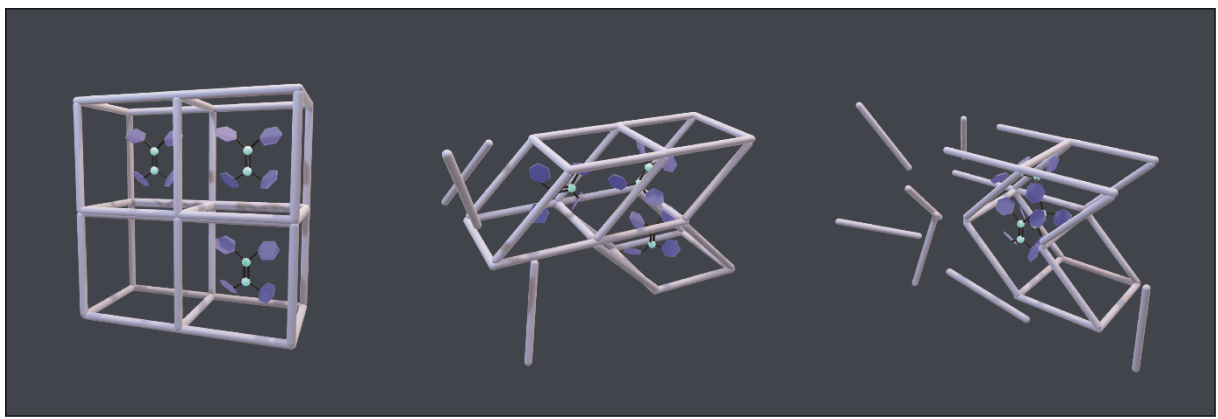

**Scheme 1.** Proposed Deformation of the TPE@ZIF-71 Structure Subject to a Mechanical Pressure. White sticks represent the secondary building units (SBUs) of ZIF-71. Purple hexagons, cyan balls, and black sticks represent the TPE molecules. As pressure increases, the ZIF-71 structure begins to deform and shear under stress, resulting in a stronger caging effect. When pressure continues to increase, some ZIF-71 pores are collapsed and amorphized, and the TPE molecules in the adjacent pores will form aggregates.

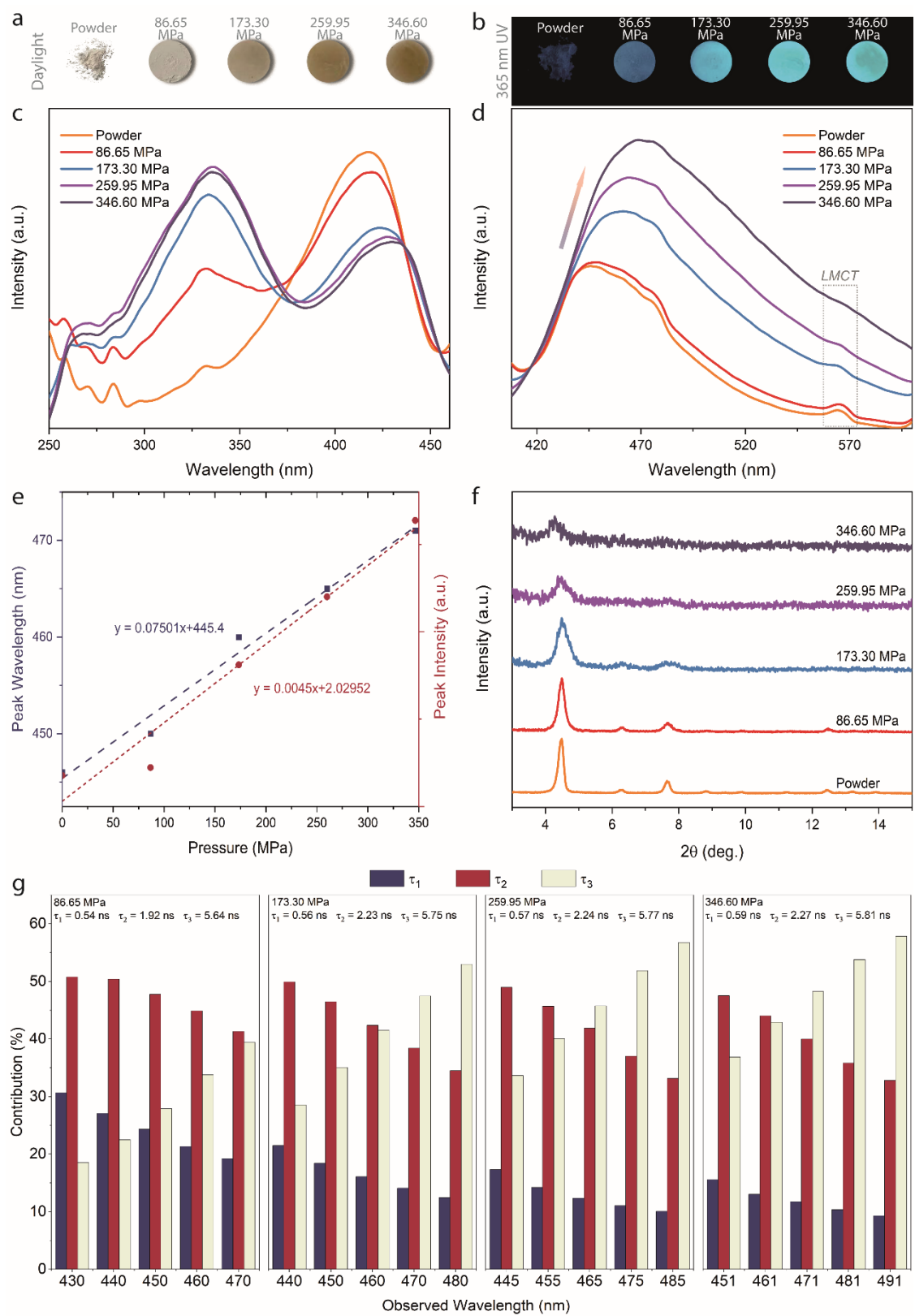

**Figure 3.** a) TPE@ZIF-71 pellets prepared under different pressures, their colors viewed in ambient light, and b) their fluorescence observed under the 365 nm UV lamp. c) Normalized excitation spectra and d) normalized emission spectra of the TPE@ZIF-71 powder and pellets.

e) Linear relationship between the emission peak wavelength, peak intensity, and the applied pelleting pressure for TPE@ZIF-71. f) PXRD patterns of the TPE@ZIF-71 powder and pellets. g) Time constants ($\tau_i$) and fractional contributions from fluorescence lifetime measurements of TPE@ZIF-71 pellets.

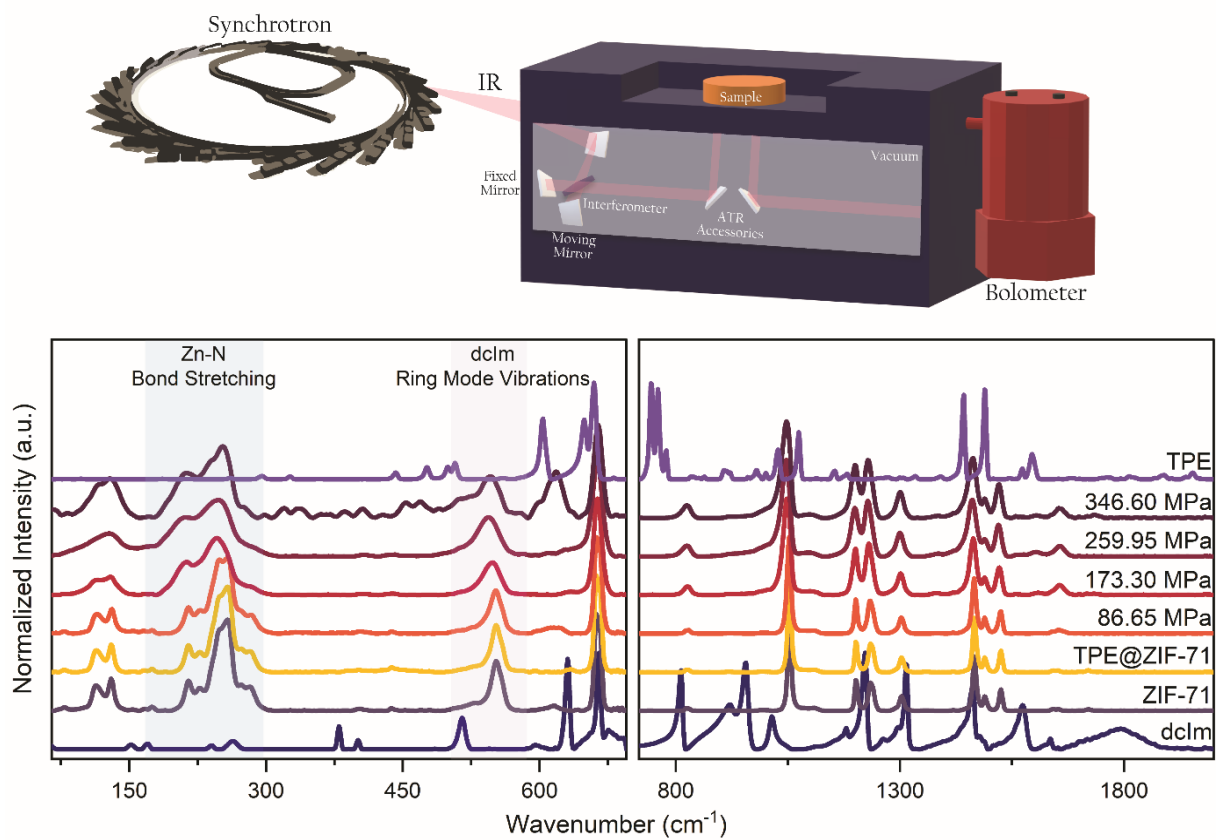

**Figure 4.** Diamond Light Source synchrotron used for SR-FTIR characterization (upper row); SR-FTIR spectra in the 65-2000 cm$^{-1}$ region (left: spectral region < 700 cm$^{-1}$ collected using a bolometer; right: > 700 cm$^{-1}$ collected from a standard detector).